\pdfoutput=1
\documentclass[aps,prd,amsmath,floats,floatfix, twocolumn,
superscriptaddress,nofootinbib,showpacs]{revtex4-1}

\usepackage[T1]{fontenc}
\usepackage[utf8]{inputenc}
\usepackage{lmodern}
\usepackage[caption=false]{subfig}
\usepackage{verbatim}

\usepackage[dvipsnames, usenames]{xcolor}
\definecolor{linkcolor}{rgb}{0.0,0.3,0.5}
\usepackage[hypertexnames=false, unicode, colorlinks=true, linkcolor=linkcolor,
citecolor=linkcolor, filecolor=linkcolor,urlcolor=linkcolor,
pdfusetitle]{hyperref}

\usepackage[all]{hypcap}
\usepackage{graphicx}
\usepackage{xspace}
\usepackage{amssymb}
\usepackage[normalem]{ulem} 
\usepackage{bm} 
\usepackage{enumitem,amssymb}
\usepackage{tikz}
\usetikzlibrary{decorations.markings}
\usetikzlibrary{calc}
\usetikzlibrary{intersections, calc}

\usepackage{microtype}
\usepackage{orcidlink}

\usepackage[english]{babel}
\usepackage{blindtext}

\graphicspath{%
  {figs/}%
}

\DeclareMathAlphabet{\mathpzc}{OT1}{pzc}{m}{it}

\newlist{todolist}{itemize}{2}
\setlist[todolist]{label=$\square$}

\begin{document}

\title{Prompt Response from Plunging Sources in Schwarzschild Spacetime}

\newcommand{\Cornell}{\affiliation{Cornell Center for Astrophysics
    and Planetary Science, Cornell University, Ithaca, New York 14853, USA}}
\newcommand\CornellPhys{\affiliation{Department of Physics, Cornell
    University, Ithaca, New York 14853, USA}}
\newcommand\Caltech{\affiliation{TAPIR 350-17, California Institute of
    Technology, 1200 E California Boulevard, Pasadena, CA 91125, USA}}
\newcommand{\AEI}{\affiliation{Max Planck Institute for Gravitational Physics
    (Albert Einstein Institute), Am M\"uhlenberg 1, Potsdam 14476, Germany}} %
\newcommand{\UMassD}{\affiliation{Department of Mathematics,
    Center for Scientific Computing and Visualization Research,
    University of Massachusetts, Dartmouth, MA 02747, USA}}
\newcommand\Olemiss{\affiliation{Department of Physics and Astronomy,
    The University of Mississippi, University, MS 38677, USA}}
\newcommand{\Bham}{\affiliation{School of Physics and Astronomy and Institute
    for Gravitational Wave Astronomy, University of Birmingham, Birmingham, B15
    2TT, UK}}
\newcommand{\Perimeter}{\affiliation{Perimeter Institute for Theoretical Physics, Waterloo, ON N2L2Y5, Canada}}

\author{Sizheng Ma\orcidlink{0000-0002-4645-453X}}
\email{sma2@perimeterinstitute.ca}
\Perimeter

\hypersetup{pdfauthor={Ma et al.}}

\date{\today}

\begin{abstract}
Gravitational waves generated by moving sources in Schwarzschild spacetime can be decomposed into three principal components: quasinormal modes, tail, and prompt response. While the first two have been extensively studied, a systematic and exact treatment of the prompt response has received comparatively little attention. In this work, building on recent progress in elucidating the structure of the Green’s function of the Regge–Wheeler equation, we place the prompt response on a firm theoretical footing and investigate its morphology for sources inspiraling and plunging into a Schwarzschild black hole. We find that during the inspiral phase, the prompt response is stronger than the dynamical excitation of quasinormal modes by a factor of $\sim$ 1.2, with both contributions modulated by the instantaneous orbital motion. Near the waveform peak, the prompt response rapidly decays, while the quasinormal modes transition into the ringdown regime. By combining the prompt response, quasinormal modes, and tail contributions, we achieve an accurate reconstruction of the full time-domain inspiral–merger–ringdown waveform at the 99$\%$ level, thereby providing strong support for the accuracy of this decomposition. These results offer new insight into the transition from inspiral to merger and ringdown.
\end{abstract}

\maketitle

\section{Introduction}
\label{sec:introduction}

Understanding the morphology of gravitational-wave (GW) waveforms emitted by merging binaries is essential for deciphering spacetime dynamics in its most dynamical and nonlinear regime.
Over the past decades, post-Newtonian theory \cite{Blanchet:2002av} and black hole (BH) perturbation theory \cite{Berti:2025hly} have been developed to describe the early inspiral \cite{Baker:2006ha,Buonanno:2006ui,Boyle:2007ft,Campanelli:2008nk,Buonanno:2009zt} and late ringdown \cite{Buonanno:2006ui,Giesler:2019uxc,Giesler:2024hcr,Baibhav:2023clw,London:2014cma} stage, thereby forming the theoretical foundation for waveform modeling and for parameterizing deviations from General Relativity \cite{Mehta:2022pcn,Agathos:2013upa,Li:2011cg,Meidam:2017dgf,Brito:2018rfr,Pompili:2025cdc,LIGOScientific:2026wpt,LIGOScientific:2026qni,LIGOScientific:2026fcf,Arun:2006hn,Will:2014kxa,Maggio:2022hre}.

Although quasinormal modes (QNMs), their nonlinear effects \cite{Mitman:2022qdl,Cheung:2022rbm,Ma:2022wpv,Khera:2023oyf,Khera:2024bjs,Ma:2024qcv,Zhu:2024rej,Redondo-Yuste:2023seq,Bucciotti:2024zyp,Bourg:2024jme,Bucciotti:2024jrv,Lagos:2024ekd,Yi:2024elj,Bourg:2025lpd,May:2024rrg,Zhu:2024dyl,Ma:2025rnv,Sberna:2021eui}, and late-time tails \cite{Price:1971fb,1972PhRvD...5.2439P} have been widely used as the primary components to model the late-time portion of waveforms and observational data (see \cite{Berti:2025hly} for a complete review), from the perspective of Green’s function decomposition, the waveform is also expected to contain a prompt response that propagates directly from the source to the observer \cite{Leaver:1986gd,Andersson:1996cm,Hui:2019aox,Lagos:2022otp,Su:2026fvj,DeAmicis:2026tus}. Despite its fundamental role, the imprint of this prompt contribution on waveforms has remained relatively unexplored. This naturally raises the question of how the prompt response manifests itself in GW signals, particularly during the inspiral–merger–ringdown transition.

Recently, a \emph{direct wave} component has been identified in both numerical-relativity and extreme-mass-ratio inspiral waveforms \cite{Oshita:2025qmn}, as well as in the analysis of the loud GW event GW250114 \cite{Lu:2025vol}, after subtracting the QNM contribution using rational filters \cite{Ma:2022wpv,Ma:2023cwe,Ma:2023vvr}.
To date, two main phenomenological methods have been proposed to model this direct wave: the steepest-descent approach \cite{Oshita:2025qmn} and the Backwards One Body model \cite{McWilliams:2018ztb,Kankani:2026byb,Kankani:2025gqj}. Both approaches are rooted in the eikonal limit, which suggests that the wave component originates from radiation emitted near the BH horizon and propagates directly to the observer. This picture consequently leads to the expectation that the direct wave is closely related to the prompt response arising in the Green’s function decomposition. It is therefore timely to establish a firm, first-principles theoretical framework for the prompt response, particularly for sources located in the near-horizon region. Such a framework would place the direct wave on a solid theoretical footing, provide new insight into the structure of GWs from binary BH mergers, and open the door to incorporating direct waves into tests of General Relativity.

Working within a high-frequency expansion, the leading contribution to the prompt response arises from the $\omega = 0$ pole of the Green’s function \cite{Andersson:1996cm}, which gives rise to a step-function behavior identical to that in flat spacetime (see, e.g., \cite{Andersson:1996cm,Hui:2019aox,Lagos:2022otp}). While higher-order contributions have recently been explored \cite{DeAmicis:2026tus} within the Mano–Suzuki–Takasugi (MST) formalism \cite{Mano:1996vt,Mano:1996mf,Sasaki:2003xr}, this treatment does not fully capture the prompt response, particularly for sources located near BHs. As we will discuss below, the missing contribution is associated with the integral along the high-frequency arc in the complex frequency plane, whose direct numerical evaluation is challenging.
Recently, it has been proposed that the high-frequency arc can be converted to a branch cut contribution \cite{Su:2026fvj} (see also \cite{Arnaudo:2025uos}), thereby rendering the previously inaccessible portion of the prompt response numerically tractable, given that extensive set of techniques have been developed for handling branch cuts \cite{Leaver:1986gd,Leung:2003ix,Casals:2011aa,Casals:2012tb,Casals:2015nja,Casals:2012ng,Casals:2013mpa}.

In this work, we demonstrate that the proposal of \cite{Su:2026fvj} can be effectively leveraged to model the prompt response from plunging sources in Schwarzschild spacetime. As a proof of principle, we adopt the Regge-Wheeler equation and consider two representative scenarios: a quasi-circular plunge from the innermost stable circular orbit (ISCO) and a radial infall. We compute the prompt response, the dynamical excitation of QNMs \cite{Andersson:1996cm,Chavda:2024awq,DeAmicis:2025xuh}, and the late-time tail separately, and show that their sum reproduces the full time-domain inspiral–merger–ringdown waveform with $\sim 99\%$ accuracy.

This paper is organized as follows. In Sec.~\ref{sec:Green_Function}, we provide an overview of the method proposed in \cite{Su:2026fvj}. In Sec.~\ref{sec:prompt_response_from_plunging_sources}, we apply this method to decompose time-domain waveforms from plunging sources. Finally, we conclude in Sec.~\ref{sec:conclusion}. Throughout this paper we set the BH mass to
$M=1$.

\section{Green’s Function of the Regge–Wheeler Equation}
\label{sec:Green_Function}
In this section, we review the structure of the Green's function associated with the Regge-Wheeler equation, following the method established in \cite{Su:2026fvj}. We focus specifically on its behavior when the source point is located in the vicinity of the BH horizon and examine how this decomposition facilitates the investigation of the prompt response in that regime.

The frequency-domain Green's function $G_{\ell}(\omega,r,r')$ satisfies
\begin{align}
    \left[ \partial_{r_*}^2+\omega^2
- V_\ell(r)
\right]
G_{\ell }(\omega,r,r^\prime)
=
\delta(r-r^\prime),
\end{align}
where $V_\ell(r)$ is the potential and $r_*$ denotes the tortoise coordinate,
\begin{align}
    r_* = r + 2 \ln\left(\frac{r}{2} - 1\right).
\end{align}
The Green's function is defined by imposing physically motivated boundary conditions: purely ingoing waves at the horizon ($r_* \to -\infty$) and purely outgoing waves at spatial infinity ($r_* \to +\infty$).

A convenient representation of $G_{\ell}$ can be constructed in terms of two linearly independent homogeneous solutions, $R^{\rm in}_\omega$ and $R^{\rm up}_\omega$, as
\begin{align}
    G_\ell(\omega,r,r^\prime)=\frac{R^{\rm up}_{\omega}(r_>)R^{\rm in}_{\omega}(r_<)}{2i\omega A^{\rm inc}_{\rm in}},
\end{align}
with $r_> \equiv \max(r,r')$ and $r_< \equiv \min(r,r')$. The homogeneous solutions are defined by their asymptotic behaviors:
\begin{align}
\label{eq:in_and_up}
    \begin{aligned}
R_{\omega}^{\rm in}(r) &\sim \begin{cases} e^{-i\omega r_{*}}, & r_{*} \rightarrow -\infty \\ A_{\rm in}^{\rm ref} e^{i\omega r_{*}} + A_{\rm in}^{\rm inc} e^{-i\omega r_{*}}, & r_{*} \rightarrow +\infty \end{cases} \\
R_{\omega}^{\rm up}(r) &\sim \begin{cases} A_{\rm up}^{\rm inc}e^{i\omega r_{*}} + A_{\rm up}^{\rm ref} e^{-i\omega r_{*}}, & r_{*} \rightarrow -\infty \\ e^{i\omega r_{*}}, & r_{*} \rightarrow +\infty \end{cases}
\end{aligned}
\end{align}
In the following, we focus on waveforms observed at infinity $(r\to \infty)$, the corresponding time-domain Green function $G_{\ell}$ reduces to
\begin{align}
    G_{\ell}(u;r^\prime)=\int_{-\infty}^\infty \frac{d\omega}{2\pi} e^{-i\omega u}  \frac{R^{\rm in}_{\omega}(r^\prime)}{2i\omega A^{\rm inc}_{\rm in}}, \label{eq:G_ell_def}
\end{align}
which depends solely on the retarded time $u=t-r_*$ for a given source location $r^\prime$.

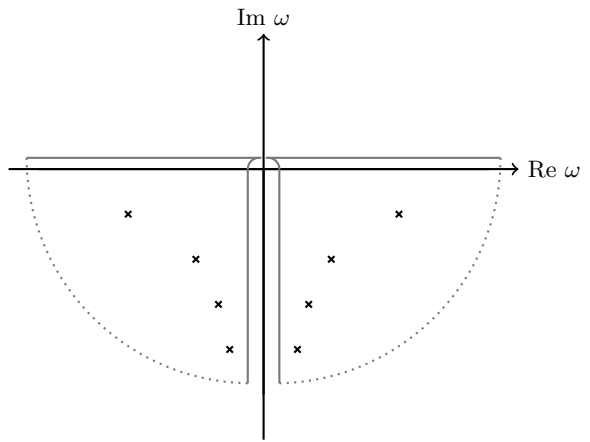
\begin{figure}[!htb]
    \hspace{0.5cm}
    \begin{tikzpicture}[scale=3]
        \draw [->, thick] (-1.13,0) -- (1.13,0) node [right] {$\rm Re$ $\omega$};
        \draw [->, thick] (0,-1.2) -- (0,0.6) node [above] {$\rm Im$ $\omega$};
        
        \draw [thick] (0,0) -- (0,-1);
        
        \draw [thick] (-1,0) -- (1,0);
        
        
        \definecolor{mygray}{RGB}{120,120,120}
        
        \draw [dotted, thick, mygray] (-1.05,0.05) arc [start angle=180, end angle=270, radius=1];
        \draw [dotted, thick, mygray] (1.05,0.05) arc [start angle=0, end angle=-90, radius=1];
        
        \draw [thick, mygray] (0.07,0.0) -- (0.07,-0.95);
        \draw [thick, mygray] (-0.07,-0.0) -- (-0.07,-0.95);
        
        \draw [thick, mygray] (-1.05,0.05) -- (-0.01,0.05);
        \draw [thick, mygray] (0.01,0.05) -- (1.05,0.05);

        \draw [thick, mygray] (0.02,0.05) arc [start angle=90, end angle=0, radius=0.05];
        \draw [thick, mygray] (-0.02,0.05) arc [start angle=90, end angle=180, radius=0.05];

        \foreach \n in {1,2,3,4} {
    \begin{scope}[shift={(0.6/\n, -0.2*\n)}, rotate=45]
        \draw[thick] (-0.02,0) -- (0.02,0);
        \draw[thick] (0,-0.02) -- (0,0.02);
    \end{scope}
    
    \begin{scope}[shift={(-0.6/\n, -0.2*\n)}, rotate=45]
        \draw[thick] (-0.02,0) -- (0.02,0);
        \draw[thick] (0,-0.02) -- (0,0.02);
    \end{scope}
}

    \end{tikzpicture}
    \caption{Integration contour for the Green's function when $u>|r_*^\prime|$. In this regime, the large-arc contribution vanishes, as indicated by the dotted curves. The Green's function is contributed by QNM poles (crosses) and the branch cut along the negative imaginary axis (solid gray lines).}
    \label{fig:GreenFunctionCount_late_time} 
\end{figure}

\subsection{Green's function decomposition}

As pointed out by Leaver \cite{Leaver:1986gd}, Eq.~\eqref{eq:G_ell_def} can be evaluated by deforming the integration contour in the complex frequency plane, as shown in Fig.~\ref{fig:GreenFunctionCount_late_time}. In this approach, the integral along the real-frequency axis is recast as the sum of three contributions: the QNM poles, the branch-cut tail integral, and the large arc.

At late times $u > |r_*^\prime|$, the large arc vanishes, and we use the dotted segment in Fig.~\ref{fig:GreenFunctionCount_late_time} to indicate this fact. The Green’s function $G_\ell(u)$ can then be decomposed as
\begin{align}
    G_\ell(u;r^\prime)=G^{\rm QNM}_\ell(u;r^\prime)+ G^{\rm Tail}_\ell(u;r^\prime).
\end{align}
The QNM contribution is obtained from the residues at the poles
\begin{align}
    G^{\rm QNM}_\ell(u;r^\prime)=-{\rm Re}\sum_{n}  e^{-i\omega_n u}\frac{R^{\rm in}_{\omega_{n}}(r^\prime)}{\omega_{n} \left.(d A^{\rm inc}/d\omega)\right|_{\omega_{n}}};  \label{eq:QNM_Green_function}
\end{align}
whereas the tail contribution arises from the integral along the negative imaginary axis \cite{Leung:2003ix,Casals:2011aa,Casals:2012tb,Casals:2015nja,Casals:2012ng,Casals:2013mpa}
\begin{align}
    G^{\rm Tail}_\ell(u;r^\prime)=-\int_{0}^{\infty} \frac{d\sigma}{2\pi} e^{-\sigma u}  \frac{R^{\rm in}_{-i\sigma}(r^\prime)}{\sigma } \frac{ {\rm Im}~A^{\rm inc}_{\rm in}}{|A^{\rm inc}_{\rm in}|^2}, \label{eq:QNM_tail_function}
\end{align}
with $\sigma=i\omega$.

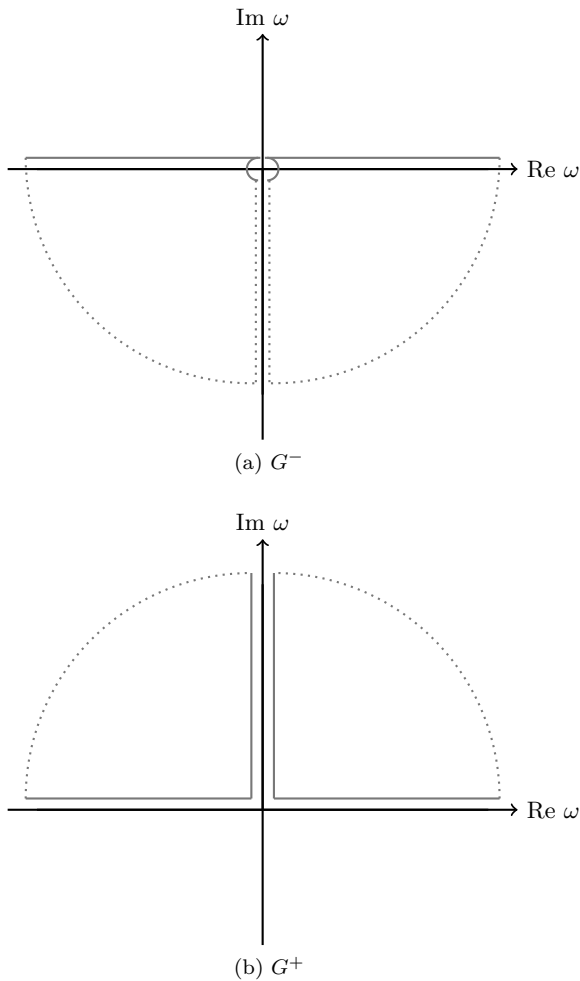
\begin{figure}[!htb]
\centering
\subfloat[$G^-$ \label{fig:Gm}]{
    \hspace{0.5cm}
    \begin{tikzpicture}[scale=3]
        \draw [->, thick] (-1.13,0) -- (1.13,0) node [right] {$\rm Re$ $\omega$};
        \draw [->, thick] (0,-1.2) -- (0,0.6) node [above] {$\rm Im$ $\omega$};
        
        \draw [thick] (0,0) -- (0,-1);
        
        \draw [thick] (-1,0) -- (1,0);
        
        
        \definecolor{mygray}{RGB}{120,120,120}
        
        \draw [dotted, thick, mygray] (-1.05,0.05) arc [start angle=180, end angle=272, radius=1];
        \draw [dotted, thick, mygray] (1.05,0.05) arc [start angle=0, end angle=-92, radius=1];
        
        \draw [dotted,thick, mygray] (0.03,-0.05) -- (0.03,-0.95);
        \draw [dotted,thick, mygray] (-0.03,-0.05) -- (-0.03,-0.95);
        
        \draw [thick, mygray] (-1.05,0.05) -- (-0.01,0.05);
        \draw [thick, mygray] (0.01,0.05) -- (1.05,0.05);

        \draw [thick, mygray] (0.02,-0.05) arc [start angle=-90, end angle=90, radius=0.05];
        \draw [thick, mygray] (-0.02,-0.05) arc [start angle=-90, end angle=-270, radius=0.05];
    \end{tikzpicture}
}\\
\subfloat[$G^+$ \label{fig:Gp}]{
\hspace{0.5cm}
\begin{tikzpicture}[scale=3]

\draw[->, thick] (-1.13,0) -- (1.13,0) node[right] {$\rm Re$ $\omega$};
\draw[->, thick] (0,-0.6) -- (0,1.2) node[above] {$\rm Im$ $\omega$};

\draw[thick] (0,0) -- (0,1);

\draw[thick] (-1,0) -- (1,0);


\definecolor{mygray}{RGB}{120,120,120}

\draw[dotted, thick, mygray] (-1.05,0.05) arc[start angle=180, end angle=90, radius=1];

\draw[dotted, thick, mygray] (1.05,0.05) arc[start angle=0, end angle=90, radius=1];

\draw[thick, mygray] (0.05,0.05) -- (0.05,1.05);
\draw[thick, mygray] (-0.05,0.05) -- (-0.05,1.05);

\draw[thick, mygray] (-1.05,0.05) -- (-0.05,0.05);
\draw[thick, mygray] (0.05,0.05) -- (1.05,0.05);

\end{tikzpicture}
}
    \caption{Integration contour for the Green's function when $-r_*^\prime<u<|r_*^\prime|$. In this regime, the full Green's function has a rather complicated structure. To facilitate the analysis, Eq.~\eqref{eq:split_green_funcition} is adopted to decompose $G_{\ell}(u;r^\prime)$ into $G^\pm$, with each admitting a simpler contour representation. The contribution $G^- (G^+)$ vanishes along the large arc in the lower (upper) plane, allowing the corresponding contour to be deformed to avoid these arcs. Consequently, $G^-$ reduces to the residue at the pole $\omega=0$, whereas $G^+$ is given by the branch cut along the positive imaginary axis. Solid (dotted) curves denote nonvanishing (vanishing) contributions.}
    \label{fig:prompt_response_green_function} 
\end{figure}

At intermediate times $-r_*^\prime<u<|r_*^\prime|$, the large-arc contribution becomes nonvanishing, and its direct evaluation is numerically challenging. As shown in \cite{Andersson:1996cm,Su:2026fvj}, this difficulty can be circumvented by first decomposing the in-solution in Eq.~\eqref{eq:G_ell_def} as
\begin{align}
    R_{\omega}^{\rm in}(r^\prime) = A_{\rm in}^{\rm inc} R_{\omega}^{\rm down}(r^\prime) + A_{\rm in}^{\rm ref} R_{\omega}^{\rm up}(r^\prime),
\end{align}
which yields
\begin{align}
    G_{\ell}(u;r^\prime)=\int_{-\infty}^\infty \frac{d\omega}{2\pi} e^{-i\omega u}  [G^-_\omega(r^\prime)+G^+_\omega(r^\prime)] , \label{eq:split_green_funcition}
\end{align}
with 
\begin{align}
     & G^-=\frac{R_{\omega}^{\rm down}(r^\prime)}{2i\omega }, &G^+=\frac{ A_{\rm in}^{\rm ref} R_{\omega}^{\rm up}(r^\prime)}{2i\omega A^{\rm inc}_{\rm in}}.
\end{align}
Here the down-solution is related to the up-solution via $R_{\omega}^{\rm down}=\bar{R}_{\bar{\omega}}^{\rm up}$.
For $G^-$, it was shown in \cite{DeAmicis:2026tus} that the contribution from the large arc in the lower half of the complex frequency plane vanishes. One may therefore close the contour as shown in Fig.~\ref{fig:Gm}, in a manner analogous to Leaver’s construction in Fig.~\ref{fig:GreenFunctionCount_late_time}. In addition, $G^-$ does not exhibit a branch cut along the negative imaginary axis  \cite{Su:2026fvj}. As a consequence, the integral along the real-frequency axis can be reduced to a contribution from a small arc around $\omega = 0$. In Fig.~\ref{fig:Gm}, the solid gray curve denotes this small-arc contribution, while the dotted segments indicate the vanishing large-arc and branch-cut contributions. In \cite{DeAmicis:2026tus}, this small-arc integral was used to model the prompt response in the low-frequency regime, particularly when the source location $r^\prime$ is far from the horizon.

However, as pointed out in \cite{Su:2026fvj} and discussed in detail below, the situation is different for $G^+$. In this case, the contribution from the large arc in the lower half of the complex frequency plane \emph{does not vanish}, and becomes increasingly important as the source location approaches the horizon. Direct numerical evaluation of $G^+$ along this large arc is again challenging.
An alternative approach is to deform the contour as shown in Fig.~\ref{fig:Gp}. With this choice, the contribution from the large arc in the upper half-plane vanishes \cite{Su:2026fvj}, and no additional (QNM) pole contributions are encountered. Consequently, $G^+$ contributes to the Green’s function entirely through the branch cut along the positive imaginary axis, and can be modeled via branch-cut techniques that have been developed in the literature \cite{Leaver:1986gd,Leung:2003ix,Casals:2011aa,Casals:2012tb,Casals:2015nja,Casals:2012ng,Casals:2013mpa}. This yields a nonnegligible contribution to the prompt response especially when $r^\prime$ is small.

Finally, for completeness, we note that the total Green’s function $G_\ell$ vanishes at early times $u < -r_*^\prime$, as required by causality.

\subsection{Numerical results}
Having summarized the general structure of the Green’s function decomposition, we now present explicit examples illustrating how $G^+$ and $G^-$ together reconstruct the prompt response within the time window $-r_*^\prime < u < |r_*^\prime|$ for a range of source locations $r^\prime$, including regions near and within the light ring.

Following \cite{Su:2026fvj}, we employ the MST formalism \cite{Mano:1996vt,Mano:1996mf,Sasaki:2003xr} to construct the up- and down-solutions, as well as the associated amplitudes in Eq.~\eqref{eq:in_and_up}, along the small arc in Fig.~\ref{fig:Gm} and the positive imaginary axis in Fig.~\ref{fig:Gp}. Implementation details can be found in \cite{Su:2026fvj}.

\begin{figure*}[!htb]
        \subfloat[$r^\prime=60$]{\includegraphics[width=\columnwidth]{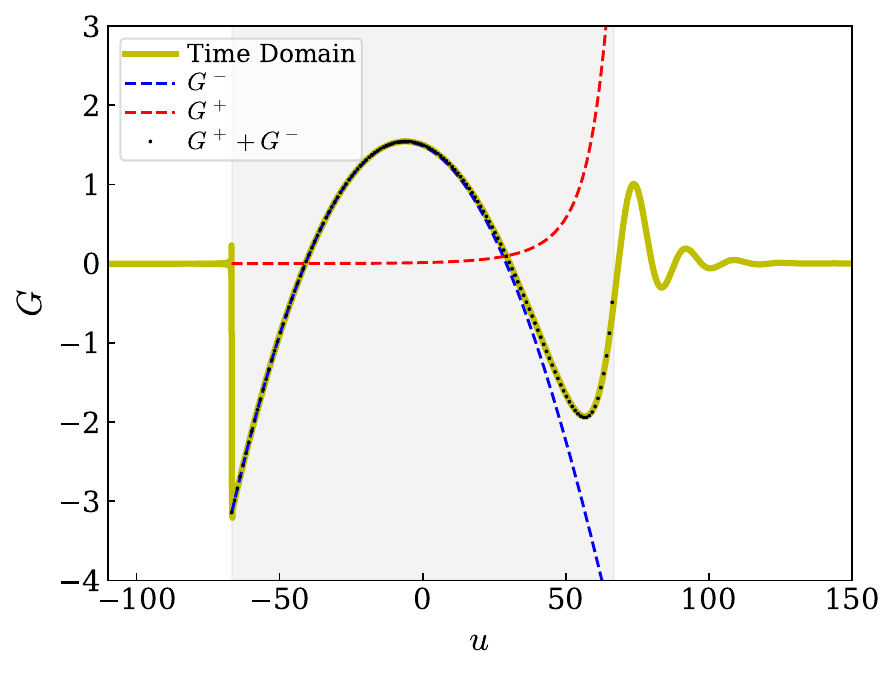}}
        \subfloat[$r^\prime=20$]{\includegraphics[width=\columnwidth]{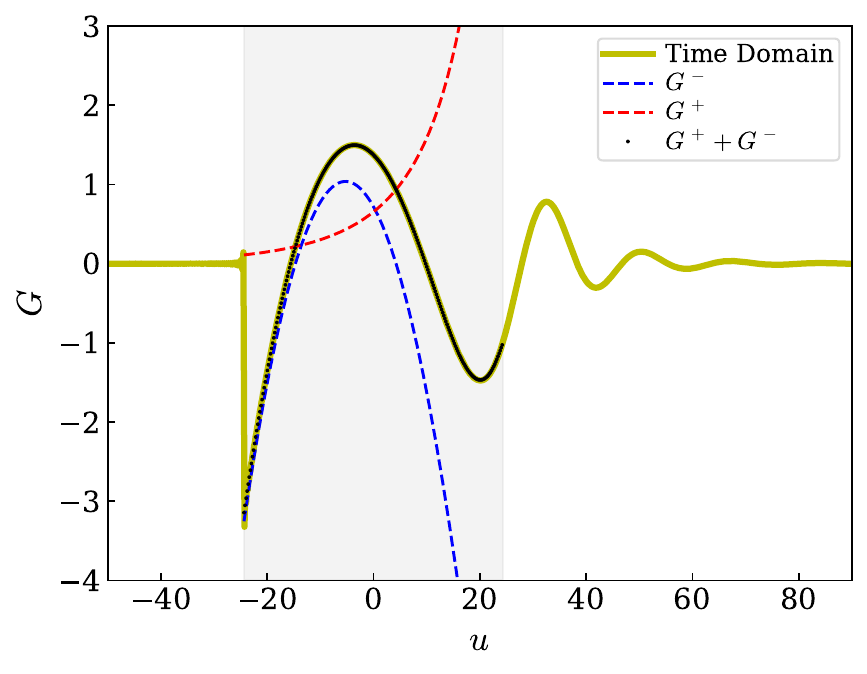}} \\
        \subfloat[$r^\prime=3$ \label{fig:GreenFunctionAtLocations:r3}]{\includegraphics[width=\columnwidth]{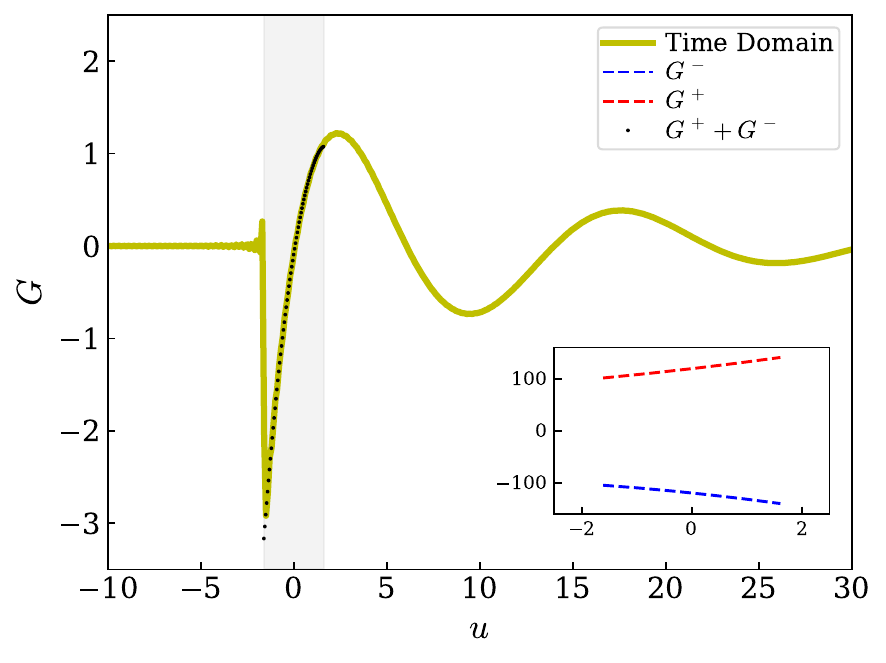}}
        \subfloat[$r^\prime=2.6$\label{fig:GreenFunctionAtLocations:r2.6}]{\includegraphics[width=\columnwidth]{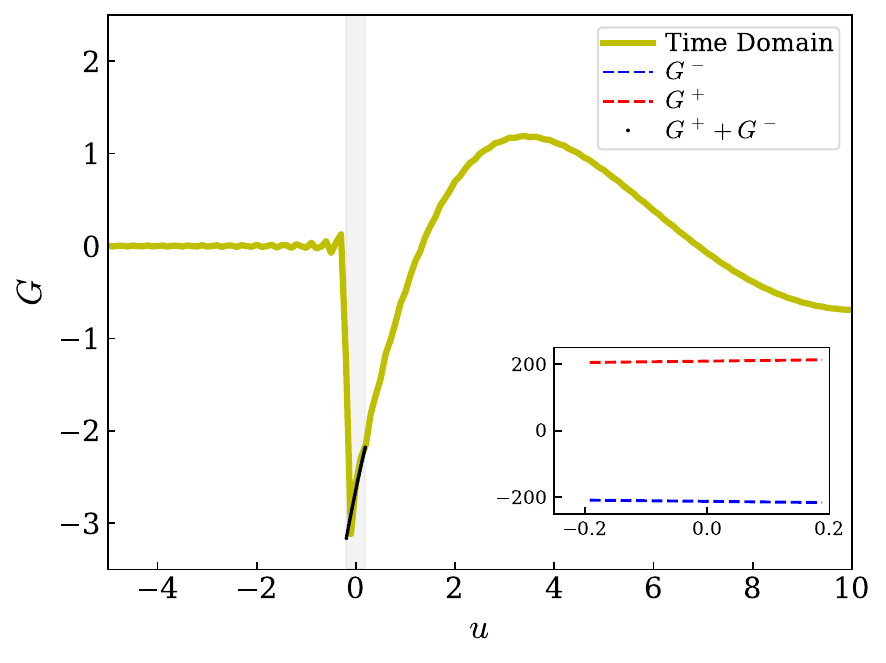}}
    \caption{Comparison of the full time-domain Green’s function (yellow) with the prompt response computed using Eq.~\eqref{eq:split_green_funcition} and Fig.~\ref{fig:prompt_response_green_function}, within the time window $-r_*^\prime<u<|r_*^\prime|$ (gray regions) for various source locations $r^\prime$. In each case, $G^{\pm}$ are shown as blue/red dashed curves, while their sum, $G^++G^-$, is indicated by the black dots. The contribution from $G^+$ (branch cut) becomes increasingly important as $r^\prime$ approaches the horizon. Near the light ring $r^\prime=3$, $G^\pm$ have large magnitudes with opposite signs (inset panels in Fig.~\ref{fig:GreenFunctionAtLocations:r3} and \ref{fig:GreenFunctionAtLocations:r2.6}). The resulting net contribution remains of order unity and continues to agree well with the full time-domain signal. }
    \label{fig:GreenFunctionAtLocations}
\end{figure*}

In Fig.~\ref{fig:GreenFunctionAtLocations}, we display the full time-domain Green's function (yellow solid curves) for source locations $r^\prime=60,20,3$ and 2.6, separately. For each case, we also show the contributions from $G^-$ (blue dashed curves) and $G^+$ (red dashed curves), together with their sum (black dotted curves), restricted to the time window $-r_*^\prime < u < |r_*^\prime|$ (shaded gray regions).

For a distant source at $r^\prime = 60$, the magnitude of $G^+$ is small compared to that of $G^-$. It becomes significant only as $u$ approaches $|r_*^\prime|$. Consequently, the small-arc contribution in Fig.~\ref{fig:Gm} provides a good approximation to the prompt response over most of the time window, consistent with the findings of \cite{DeAmicis:2026tus}, and also previous leading order calculations \cite{Andersson:1996cm,Hui:2019aox,Lagos:2022otp}. Meanwhile, the combined contribution $G^+ + G^-$ accurately reproduces the full time-domain signal throughout the entire interval $-r_*^\prime < u < |r_*^\prime|$.

As the source approaches the BH, the contribution from $G^+$ becomes increasingly important. In particular, near the light ring ($r \sim 3$), $G^+$ and $G^-$ attain comparable magnitudes ($\sim 100$–$200$) but with opposite signs. This behavior is evident in the inset panels of Figs.~\ref{fig:GreenFunctionAtLocations:r3} and \ref{fig:GreenFunctionAtLocations:r2.6}, where the two contributions largely cancel, yielding a net amplitude of order unity ($\sim 1$). Nevertheless, the combined contribution $G^+ + G^-$ continues to accurately reproduce the prompt response within the relevant time window, even when the source lies within the light ring.

Since the prompt response is confined to the interval $-r_*^\prime<u<|r_*^\prime|$, the width of this time window, $|r_*^\prime| + r_*^\prime$, decreases as $r_*^\prime$ decreases, and vanishes once $r_*^\prime$ becomes negative. This occurs at $r \leq 2[1 + W(1/e)] \approx 2.5569$, where $W$ denotes the Lambert $W$ function. In this regime, the prompt response is localized to the first time step of Green's function, while the subsequent evolution is fully described by the QNM and tail contributions given in Eqs.~\eqref{eq:QNM_Green_function} and \eqref{eq:QNM_tail_function}.

\begin{figure}[!htb]
    \includegraphics[width=\columnwidth]{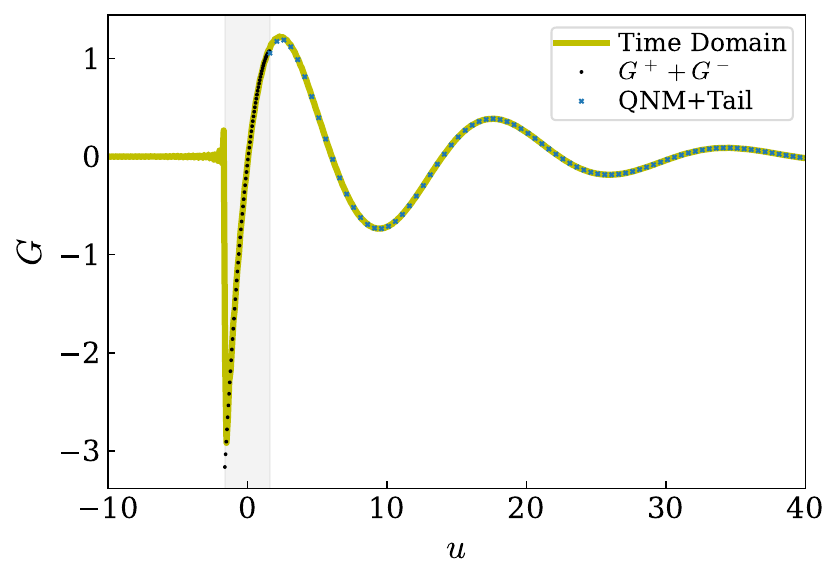}
    \caption{Time-domain Green’s function (yellow) generated by a source at the light ring $r^\prime=3$. The prompt response (black dots), computed via Eq.~\eqref{eq:split_green_funcition} and Fig.~\ref{fig:prompt_response_green_function}, is shown within the time window $-r_*^\prime<u<|r_*^\prime|$ (gray region), while the QNM+tail contribution (blue crosses), computed using Eqs.~\eqref{eq:QNM_Green_function} and \eqref{eq:QNM_tail_function}, is shown for $u>|r_*^\prime|$. The two pieces join smoothly at the boundary $u=|r_*^\prime|$.  }
    \label{fig:GreenFunction_at3_with_QNM} 
\end{figure}

For completeness, we close this section by noting that the Green’s function for $u > |r_*^\prime|$ is well described by QNMs and the tail, as extensively studied in the literature \cite{Berti:2025hly}. Figure~\ref{fig:GreenFunction_at3_with_QNM} illustrates this behavior for a source at $r^\prime = 3$. The prompt-response and QNM+tail regimes join smoothly, with no discontinuity at the transition time $u = |r_*^\prime|$. This provides a theoretical foundation for decomposing an inspiral–merger–ringdown waveform into prompt-response, QNM, and tail contributions, via
\begin{align}
    G_{\ell}(u;r^\prime)=\Theta(u+r_*^\prime)&\left[G^{\rm Prompt}_{\ell}(u;r^\prime)\Theta(|r_*^\prime|-u) \right. \notag \\
    &\left.+G^{\rm QNM+Tail}_{\ell}(u;r^\prime)\Theta(u-|r_*^\prime|)\right], \label{eq:GreenDecompTimeDomain}
\end{align}
where $\Theta$ is the Heaviside step function.

\section{Prompt response from plunging Sources}
\label{sec:prompt_response_from_plunging_sources}
In the presence of a source, the solution to the Regge-Wheeler equation,
\begin{align}
    \left[-\partial_t^2
+ \partial_{r_*}^2
- V_\ell(r)
\right]
\Psi_{\ell }(t,r)
=
S(t,r),
\end{align}
can be written as
\begin{align}
    \Psi_{\ell}(t,r)=\int_{-\infty}^\infty \frac{d\omega}{2\pi} e^{-i\omega t} \int dr^\prime G_\ell(\omega,r,r^\prime) S(\omega,r^\prime),
\end{align}
with the source in the frequency domain given by
\begin{align}
    S(\omega,r)&= \int_{-\infty}^{\infty} dt\, e^{i\omega t} S(t,r).
\end{align}
We now consider a pointlike source of the form 
\begin{align}
    S(t,r) = F(t)\delta\bigl(r - r_p(t)\bigr), \label{eq:S(t,r)}
\end{align}
where $r_p(t)$ describes the radial motion of the source.
In this case, the solution at future null infinity can be expressed as
\begin{align}
    \Psi_{\ell}(u)=\int_{-\infty}^{\infty} dt^\prime\, G_{\ell}\bigl(u - t^\prime; r_p(t^\prime)\bigr) F(t^\prime),
\end{align}
where $G_{\ell}(u-t^\prime;r_p(t^\prime))$ is defined in Eq.~\eqref{eq:G_ell_def}.
Utilizing the decomposition in Eq.~\eqref{eq:GreenDecompTimeDomain}, the solution can be separated as 
\begin{align}
    \Psi_{\ell}(u)=\Psi^{\rm Prompt}_{\ell}(u)+\Psi^{\rm QNM+Tail}_{\ell}(u),
\end{align}
with
\begin{subequations}
\label{eq:decomposition_signal_at_scri}
    \begin{align}
    &\Psi^{\rm Prompt}_{\ell}(u)=\int_{t_1}^{t_2} dt^\prime\, G^{\rm Prompt}_{\ell}\bigl(u - t^\prime; r_p(t^\prime)\bigr) F(t^\prime), \\
    &\Psi^{\rm QNM+Tail}_{\ell}(u)=\int_{-\infty}^{t_1} dt^\prime\, G^{\rm QNM+Tail}_{\ell}\bigl(u - t^\prime; r_p(t^\prime)\bigr) F(t^\prime), \label{eq:decomposition_signal_at_scri_QNM_tail}
\end{align}
\end{subequations}
where $t_{1,2}$ are defined implicitly by
\begin{align}
    & u=t_2-r_{p,*}(t_2), && u=t_1+|r_{p,*}(t_1)|. \label{eq:t1_t2_def}
\end{align}
Here $r_{p,*}$ denotes the tortoise coordinate of $r_p$. 

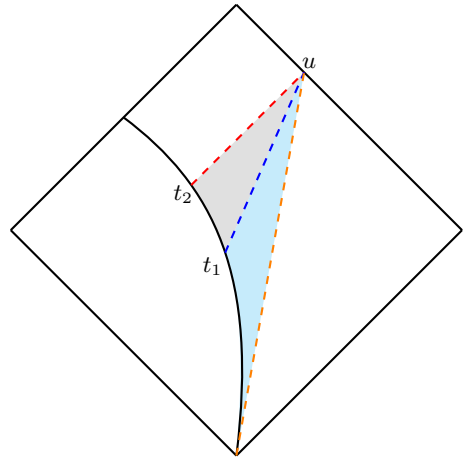
\begin{figure}[!htb]
\begin{tikzpicture}[scale=3, thick]

\draw (-1,0) -- (0,1);
\draw (0,1)  -- (1,0);
\draw (1,0)  -- (0,-1);
\draw (0,-1) -- (-1,0);

\coordinate (A) at (-0.2,0.2);      
\coordinate (B) at (0.3,0.7);       
\coordinate (C) at (-0.05,-0.1);    

\fill [gray!40, opacity=0.6] 
  (A) -- (B) -- (C) 
  .. controls (-0.08, -0.05) and (-0.12, 0.1) .. (A) -- cycle;
\fill [cyan!40, opacity=0.5] 
  (B) -- (0, -1) 
  .. controls (0.05, -0.4) and (-0.02, -0.15) .. (C) 
  -- (B) -- cycle;

\draw (-0.5, 0.5)
  .. controls (-0.1, 0.2) and (0.1, -0.2) ..
  (0, -1);

\draw[red, dashed] (A) -- (B);
\draw[blue, dashed] (C) -- (B);
\draw[orange, dashed] (0, -1) -- (B);

\node at (0.25, 0.68) [above right] {$u$};
\node at (-0.15, 0.24) [below left] {$t_2$};
\node at (-0.02, -0.08) [below left] {$t_1$};

\end{tikzpicture}
    \caption{Penrose diagram illustrating the signal decomposition for a particle plunging into the future horizon along a trajectory (black curve). The waveform observed at future null infinity at retarded time $u$ is contributed by a prompt response sourced within the time interval $[t_1, t_2]$ (gray region), and a QNM+tail component sourced from the earlier portion $(-\infty, t_1]$ (cyan region). The times $t_{1,2}$ are defined implicitly in Eq.~\eqref{eq:t1_t2_def}.  }
    \label{fig:penrose_diagram_plunging_source} 
\end{figure}

The physical interpretation of this decomposition is straightforward. As illustrated in the Penrose diagram in Fig.~\ref{fig:penrose_diagram_plunging_source}, the signal measured at future null infinity at retarded time $u$ receives contributions from two distinct intervals: a prompt-response component emitted within the time window $[t_1, t_2]$, and a QNM+tail component sourced from the earlier interval $(-\infty, t_1]$. Repeating this construction for each value of $u$ yields a decomposition of the full waveform $\Psi_\ell(u)$ into prompt-response and QNM+tail contributions.

In the following, we consider two scenarios to illustrate this decomposition.

\begin{figure*}[!htb]
    \includegraphics[width=\columnwidth]{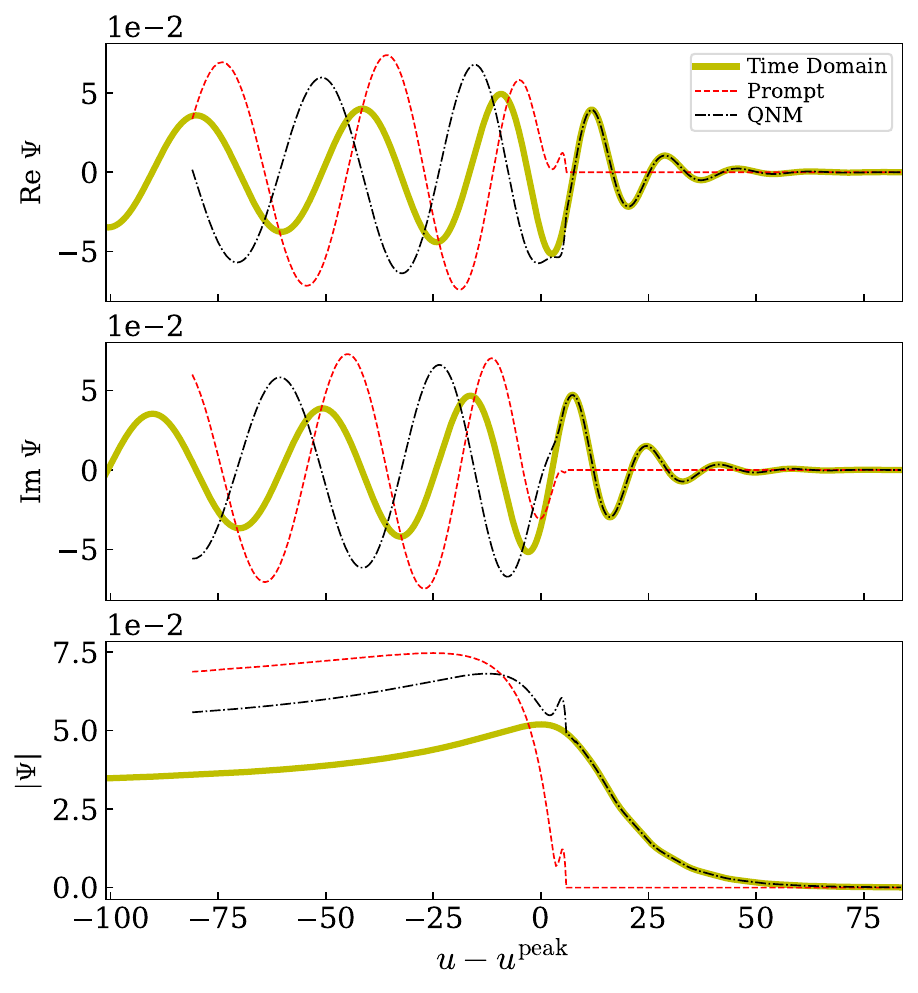}
    \includegraphics[width=\columnwidth]{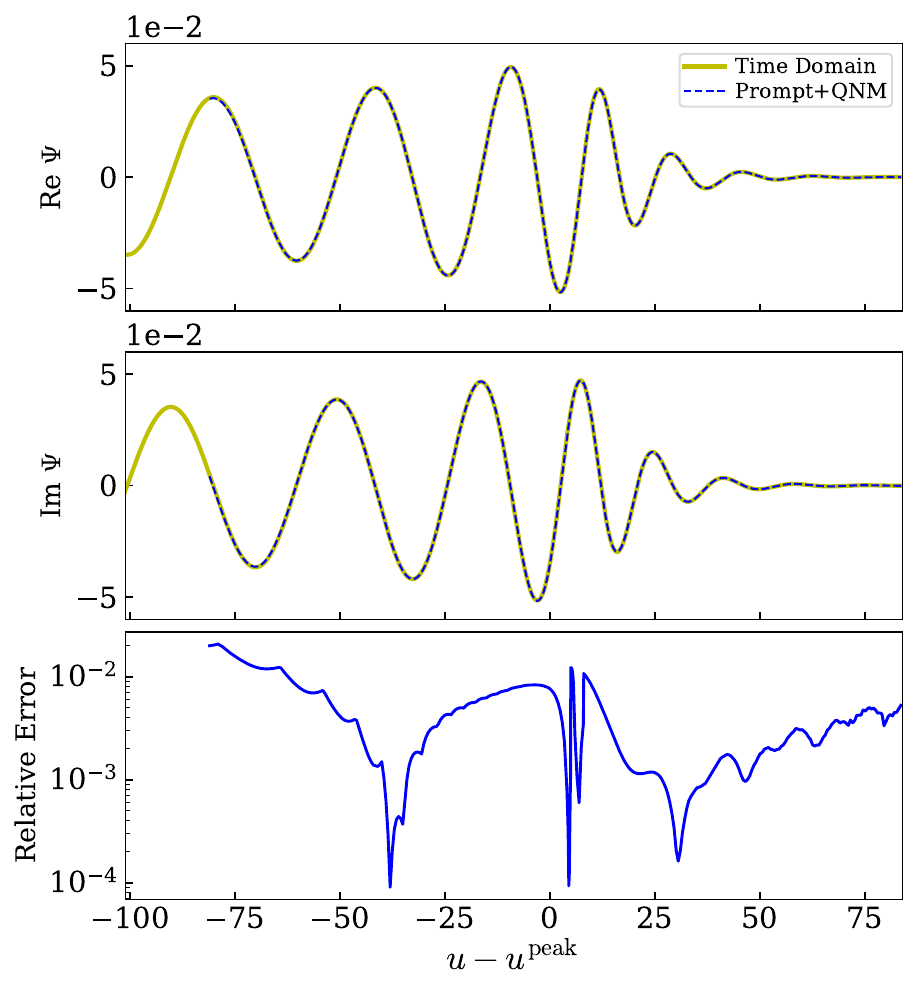}
    \caption{Waveforms emitted by a particle plunging from ISCO into the BH horizon, whose trajectory follows Eq.~\eqref{eq:ISCO_geodesic}. Left panels: The yellow curves correspond to the real (top), imaginary (middle), and absolute value (bottom) of the full time-domain waveform. The prompt response (red dashed) and QNM (black dot-dashed) contributions are computed separately via the integration in Eq.~\eqref{eq:decomposition_signal_at_scri}, with the corresponding Green's functions established in Sec.~\ref{sec:Green_Function}. For the QNM contribution, including overtones up to $n=5$ is sufficient to accurately capture the time-domain signal. Right panels: Comparison of the sum $\Psi^{\rm Prompt} + \Psi^{\rm QNM}$ (blue dashed) with the full time-domain signal (yellow solid) for the real (top) and imaginary (middle) parts. The bottom panel shows the relative error, which is below $2\%$. }
    \label{fig:ISCO_comparison_PQT} 
\end{figure*}

\subsection{Plunging from the innermost stable circular orbit }
We first consider a particle plunging from ISCO, whose trajectory is governed by the geodesic equations
\begin{subequations}
\label{eq:ISCO_geodesic}
    \begin{align}
&\frac{dt_p}{d\tau} =  \frac{1}{r_p-2} E r_p\,,\\
&\frac{dr_p}{d\tau}=-\frac{1}{3}\left(\frac{6}{r_p}-1\right)^{3/2},\\
&r_p^2 \frac{d\phi_p}{d\tau} = L_z,
\end{align}
\end{subequations}
where $\tau$ is the proper time, and $E = 2\sqrt{2}/3$ and $L_z = 2\sqrt{3}$ are the conserved energy and angular momentum, respectively. Integrating these equations yields
\begin{subequations}
\label{eq:ISCO_plunge_integration_result}
    \begin{align}
        t_p&=\frac{2(r_p-24)}{\sqrt{\frac{3}{r_p} - \frac{1}{2}}}
- 44 \sqrt{2}\,\arctan\!\left(\sqrt{\frac{6}{r_p} - 1}\right) \notag \\
&+ 4\,\operatorname{arctanh}\!\left(\sqrt{\frac{3}{r_p} - \frac{1}{2}}\right),\\
\phi_p&=-\frac{2}{\sqrt{\frac{2}{r_p} - \frac{1}{3}}}.
    \end{align}
\end{subequations}
As a proof of principle, we adopt a specific form for $F$ in Eq.~\eqref{eq:S(t,r)},
\begin{align}
F = \frac{(r_p^2 - 2r_p)^2}{r_p^6} e^{-i m \phi_p},
\label{eq:ISCO_plunging_F}
\end{align}
and note that our results are expected to remain qualitatively similar for other source prescriptions. In Eq.~\eqref{eq:ISCO_plunging_F}, $m$ denotes the azimuthal index, and we take $m = 2$.

The left panel of Fig.~\ref{fig:ISCO_comparison_PQT} shows the real (top), imaginary (middle), and absolute (bottom) values of the full time-domain signal $\Psi(u)$ (yellow curves). The transition from late inspiral to merger and ringdown is clearly visible.
We then substitute Eqs.~\eqref{eq:ISCO_plunge_integration_result} and \eqref{eq:ISCO_plunging_F} into Eq.~\eqref{eq:decomposition_signal_at_scri} to evaluate the prompt, QNM, and tail contributions, where the construction of the corresponding Green’s functions $G_\ell^{\rm Prompt/QNM/Tail}$ is described in Sec.~\ref{sec:Green_Function}.

The red curves in the left panels of Fig.~\ref{fig:ISCO_comparison_PQT} show the time evolution of $\Psi^{\rm Prompt}(u)$. During the late-inspiral stage, the signal is modulated by the instantaneous orbital motion of the particle and oscillates at twice the orbital frequency. Near the waveform peak, the magnitude of the prompt response decreases sharply (bottom left panel of Fig.~\ref{fig:ISCO_comparison_PQT}). This behavior arises from two effects. First, as the particle approaches the horizon, the time window for the prompt response, $[t_1, t_2]$ (see Fig.~\ref{fig:penrose_diagram_plunging_source}), progressively shrinks and ultimately vanishes in the near-horizon region, as discussed in Sec.~\ref{sec:Green_Function}. Second, the source term $F$ considered in Eq.~\eqref{eq:ISCO_plunging_F} approaches 0 as $r_p\to 2$, reflecting the redshift effect.

For the QNM+tail contribution, we find that the tail component is negligible. By combining Eq.~\eqref{eq:QNM_Green_function} with Eq.~\eqref{eq:decomposition_signal_at_scri}, the QNM contribution can be formally written as
\begin{align}
    \Psi^{\rm QNM}(u) = \sum_n C_n(u)\, e^{-i\omega_n u},
    \label{eq:QNM_dynamical_excitation_activation}
\end{align}
which is valid at any retarded time $u$. During the inspiral stage, the mode amplitudes $C_n(u)$ are time dependent, reflecting the dynamical excitation of QNMs \cite{Andersson:1996cm,Chavda:2024awq,DeAmicis:2025xuh}.

\begin{figure}[!htb]
    \includegraphics[width=\columnwidth]{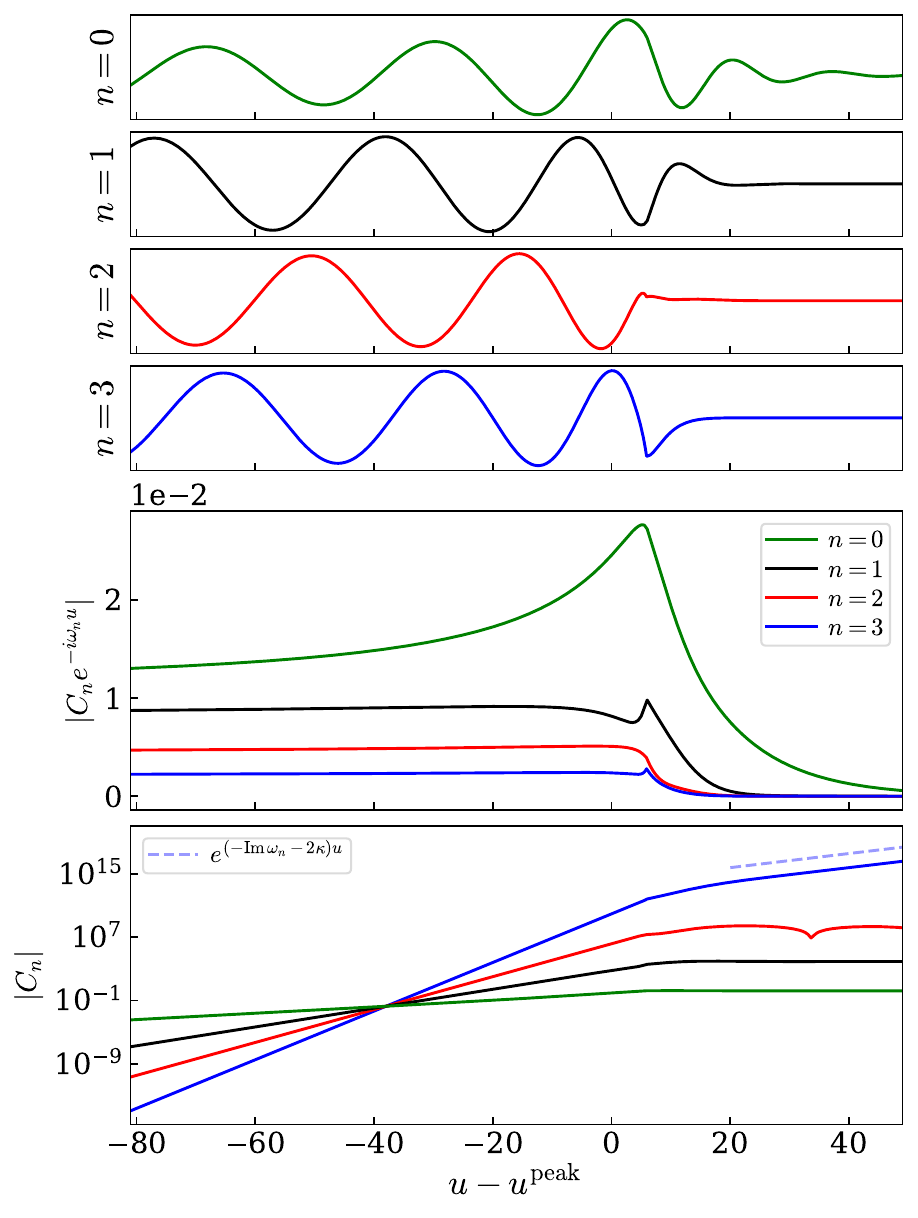}
    \caption{Dynamical excitation of QNMs, as given in Eq.~\eqref{eq:QNM_dynamical_excitation_activation}, for a particle plunging from the ISCO. The first four rows show the real part of $C_n(u)\, e^{-i\omega_n u}$ for $n=0\ldots3$, while the fifth row displays their absolute values. The last row presents the time evolution of $|C_n(u)|$.  }
    \label{fig:QNM_dynamical_excitation} 
\end{figure}

The time evolution of $\Psi^{\rm QNM}(u)$ is shown as black dashed curves in the left panels of Fig.~\ref{fig:ISCO_comparison_PQT}. Similar to the prompt response, in the early inspiral stage $\Psi^{\rm QNM}(u)$ is modulated by the instantaneous orbital motion of the particle. Its amplitude is smaller than that of the prompt contribution by a factor of $\sim 1.2$. In addition, $\Psi^{\rm Prompt}$ and $\Psi^{\rm QNM}$ differ by a phase of $\sim 1.17\pi$, leading to partial destructive interference in the full time-domain signal $\Psi(u)$. 
As the system approaches the waveform peak, $\Psi^{\rm QNM}$ grows in relative importance while $\Psi^{\rm Prompt}$ rapidly decays. Consequently, $\Psi^{\rm QNM}$ becomes the dominant contribution and transitions into the ringdown regime.

In the right panels of Fig.~\ref{fig:ISCO_comparison_PQT}, we assess the accuracy of our calculation by comparing the sum $\Psi^{\rm Prompt} + \Psi^{\rm QNM}$ against the full time-domain signal. The real and imaginary parts of the reconstructed signal are shown as blue dashed curves in the top and middle panels, respectively. We find that the relative error between $\Psi^{\rm Prompt} + \Psi^{\rm QNM}$ and $\Psi(u)$ remains at the level of $\sim 2\%$ throughout the evolution. For the QNM contribution, we include overtones up to $n=5$ in Eq.~\eqref{eq:QNM_dynamical_excitation_activation}, and we have verified that adding more overtones does not improve the accuracy.

For completeness, we examine the dynamical excitation of QNMs in Fig.~\ref{fig:QNM_dynamical_excitation}. The first four rows display the evolution of the real part of $C_n(u)\, e^{-i\omega_n u}$ for $n=0\ldots3$, while the fifth row shows their absolute values. During the early inspiral stage, the amplitudes of the overtones remain approximately constant, whereas the fundamental mode exhibits a chirping feature. The phases track the orbital evolution and alternate in sign between successive overtones\footnote{This is qualitatively consistent with discussions in \cite{McWilliams:2018ztb,Ma:2021znq}.}.
The time evolution of $C_n(u)$ is shown in the last row, where the coefficients grow exponentially  during the inspiral stage at rates set by the corresponding QNM decay rates. This exponential growth is compensated by the decaying factor $e^{-i\omega_n u}$, resulting in an overall approximately constant amplitude. Interestingly, in the present case, all $C_n(u)$ intersect at $u\sim-38$.

During the late ringdown stage, $C_{0,1}(u)$ gradually settle to constant values, whereas $C_{3}(u)$ continues to grow exponentially. This is because the source near the horizon decays as $e^{2\kappa r^\prime_*}$ when it asymptotes to the horizon, where $\kappa=1/4$ is the surface gravity. On the other hand, the QNM eigenfunctions grow as $e^{-i\omega_n r_*}$. For low $n$, the decay of the source dominates over the growth of the QNM eigenfunctions, causing the integral in Eq.~\eqref{eq:decomposition_signal_at_scri} to converge to a constant. However, starting from $n=3$, the QNM growth dominates, and the integral instead follows $e^{(-\text{Im }\omega_n-2\kappa)u}$. This behavior is illustrated by the blue dashed line in the bottom panel of Fig.~\ref{fig:QNM_dynamical_excitation}.

\subsection{Radial infall }
In the second scenario, we investigate a particle falling radially into the horizon. Its trajectory is described by
\begin{subequations}
\label{eq:radial_infall_geodesic}
    \begin{align}
 \frac{dt_p}{d\tau} &=  \frac{E r_p }{r_p-2} \,,\\
r_p \frac{dr_p}{d\tau} &= -\sqrt{E^2r_p^2 -r_p(r_p-2)}\,,
\end{align}
\end{subequations}
where we set $E=1$. Integrating the equations yields
\begin{align}
    t_p=-\frac{\sqrt{2r_p}}{3}(r_p+6)+4 \tanh^{-1}\sqrt{\frac{2}{r_p}}. \label{eq:tp_vs_rp_radial_infall}
\end{align}

\begin{figure}[!htb]
    \includegraphics[width=\columnwidth]{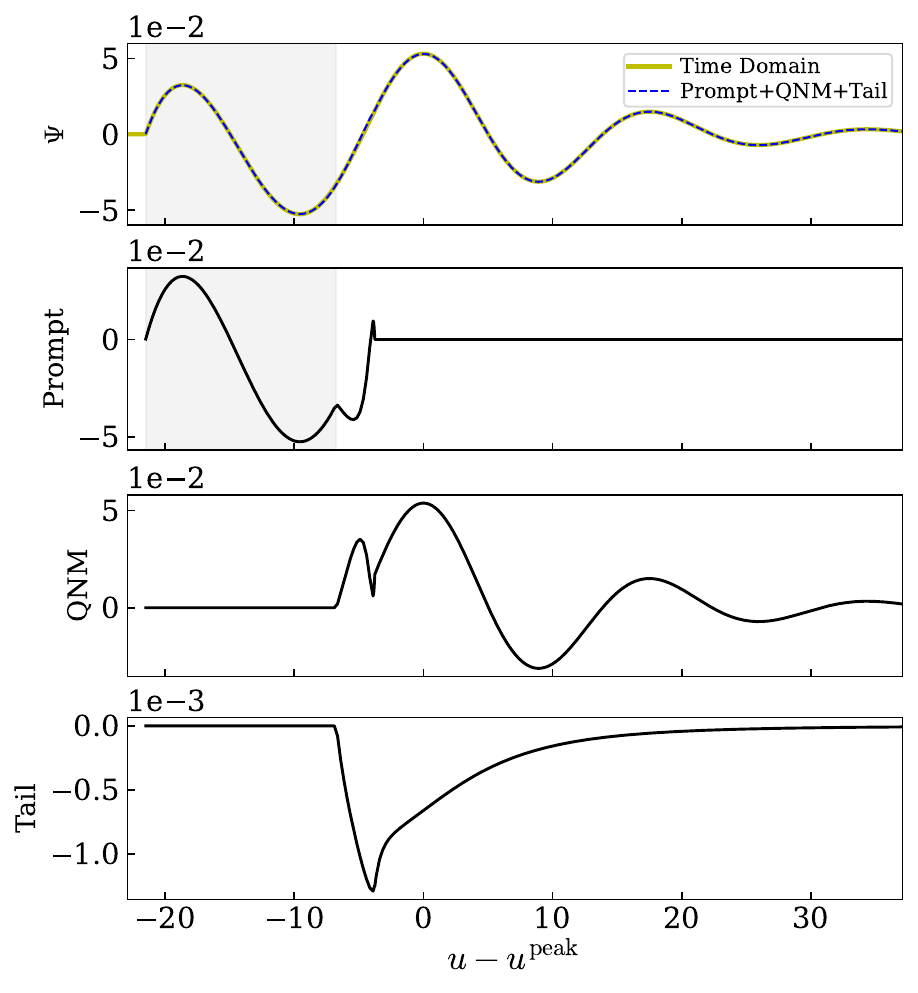}
    \caption{Waveforms emitted by a particle undergoing radial infall, following the trajectory in Eq.~\eqref{eq:radial_infall_geodesic}. The yellow curve in the top panel shows the full time-domain signal, which is compared with the reconstructed waveform obtained by summing the prompt response, QNMs, and tail contributions (blue dashed). The individual components are displayed in the subsequent rows. The gray region marks an early-time window during which only the prompt response contributes to the signal.}
    \label{fig:radial_comparison_PQT} 
\end{figure}

To study the near-horizon behavior, we initialize the particle at $r_p=6$. We adopt the same source term $F$ as in Eq.~\eqref{eq:ISCO_plunging_F} for Eq.~\eqref{eq:S(t,r)}, and focus on the $m=0$ mode. 
The resulting time-domain waveform is real-valued, whose evolution is shown by the yellow curve in the top panel of Fig.~\ref{fig:radial_comparison_PQT}.
We again compute contributions from the prompt response (second row), QNMs (third row), and tail (last row). Their sum (blue dashed curve in the top panel) closely reproduces the full signal, with error $<10^{-3}$.

Since the particle is initialized at $r_p=6$, the waveform does not immediately receive contributions from QNMs and the tail, as these backscattering effects require a finite propagation time. Consequently, there exists an early-time window during which the signal is entirely dominated by the prompt response.
From a mathematical perspective, the integral used to evaluate the QNM+tail contributions in Eq.~\eqref{eq:decomposition_signal_at_scri_QNM_tail} should begin at $t_p({r_p=6})$, as given by Eq.~\eqref{eq:tp_vs_rp_radial_infall}, rather than at $-\infty$. As a result, the integration window remains zero whenever $t_1$ (computed from Eq.~\eqref{eq:t1_t2_def} for a given $u$) lies earlier than $t_p({r_p=6})$. In the present case, the earliest such retarded time is $-6.7362$. In Fig.~\ref{fig:radial_comparison_PQT}, this prompt-response-only interval is indicated by the gray region, within which the QNM and tail contributions vanish and the signal is entirely determined by the prompt response.
Immediately after this window, both QNM and tail contributions are excited. The prompt response and QNM components each develop a transient feature with opposite signs. The prompt contribution then rapidly decays to zero at $u\sim-3.76$, corresponding to the particle crossing $r\sim 2.5569$, while the QNM component gradually transitions into the ringdown regime. The tail contribution, which also has the opposite sign to the QNM component, is about an order of magnitude smaller. Following its excitation, the tail initially exhibits a sharp decrease before transitioning to a power-law decay.

\begin{figure}[!htb]
    \includegraphics[width=\columnwidth]{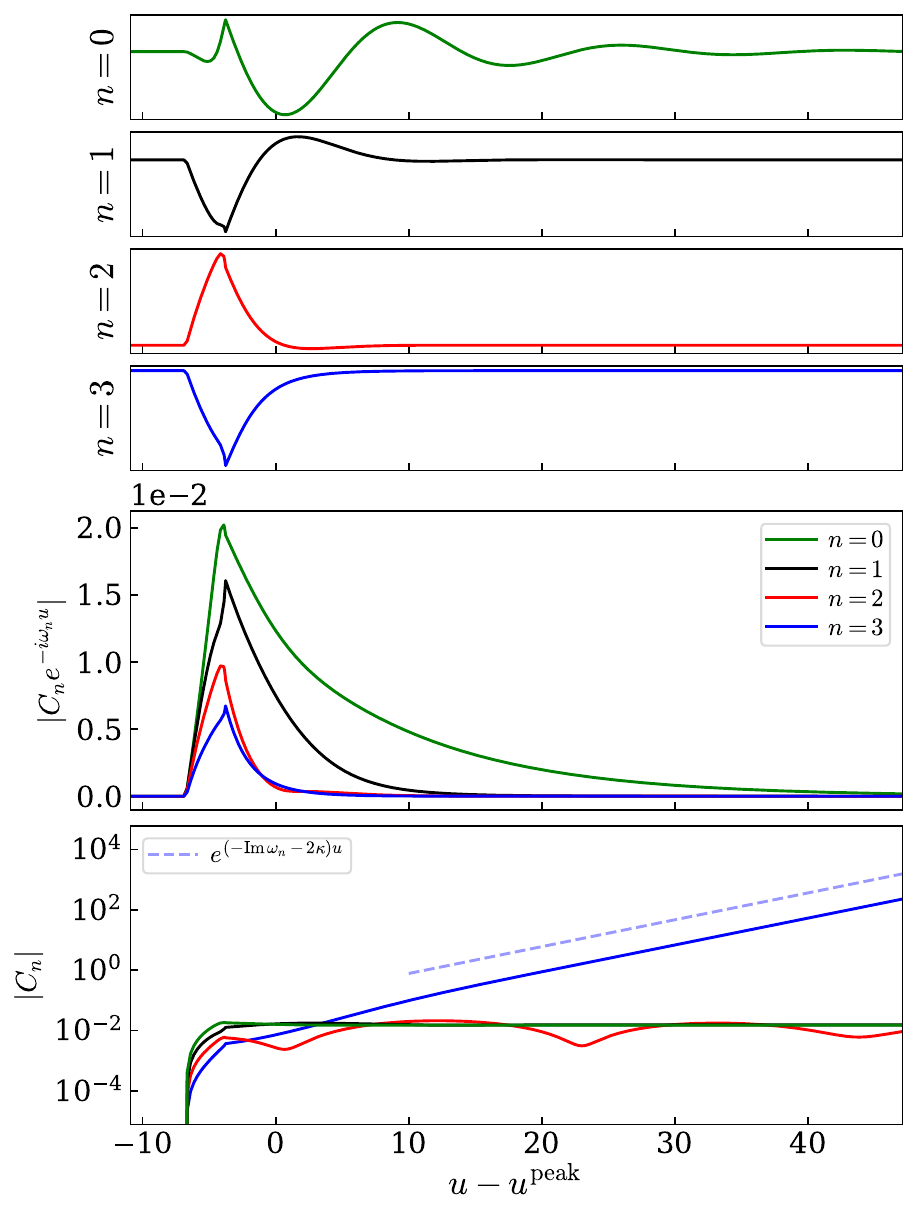}
    \caption{Similar to Fig.~\ref{fig:QNM_dynamical_excitation}. Dynamical excitation of QNMs for a particle in radial infall. }
    \label{fig:radial_QNM_dynamical_excitation} 
\end{figure}

Finally, we present the dynamical excitation of QNMs in Fig.~\ref{fig:radial_QNM_dynamical_excitation}. Each mode $|C_ne^{-i\omega_nt}|$ reaches its peak at approximately the same time that the prompt response vanishes ($u\sim -3.76$), after which $|C_n|$ approaches a constant for the fundamental mode and the first overtone. In contrast, $C_3(u)$ continues to exhibit exponential growth at late times, similar to the behavior observed in Fig.~\ref{fig:QNM_dynamical_excitation}. This arises for the same reason: the exponential growth of the QNM eigenfunction dominates over the decay of the source term near the horizon. Indeed, the growth rate is consistent with $-\text{Im }\omega_n-2\kappa$.

\section{Conclusion}
\label{sec:conclusion}
In this paper, we have computed the prompt response for plunging sources in Schwarzschild spacetime in a fully controlled manner, using the Green’s function decomposition developed in \cite{Su:2026fvj}. We showed that the prompt response arises from two distinct contributions: the pole of $G^-$ at $\omega=0$, which dominates when the source is far from the horizon, and the branch cut along the positive imaginary axis in $G^+$, which becomes increasingly important as the source approaches the BH. For near-horizon sources, the contributions from $G^\pm$ largely cancel, resulting in a net prompt response with small magnitude. At the level of the Green’s function, this decomposition provides an unambiguous definition of the temporal boundary between the prompt-response and QNM+tail regimes, with a smooth transition between the two regardless of the source's location.

Building on this theoretical foundation, we formulated a decomposition of time-domain waveforms into prompt response, QNMs, and tail contributions. Focusing on two representative scenarios: a plunge from the ISCO and a radial infall, we demonstrated that the sum of these components accurately reconstructs the full waveform. During the late inspiral stage, both the prompt response and QNM contributions are modulated by the orbital motion and are out of phase, leading to partial destructive interference. In this regime, the prompt response exceeds the QNM contribution by a factor of $\sim1.2$. Near the waveform peak, the prompt response rapidly decays, while the QNM component transitions to the ringdown regime.

To date, most efforts to understand the merger–ringdown transition and to determine the onset of ringdown have relied on QNM fitting (see \cite{Berti:2025hly} for a review) or rational filters \cite{Ma:2022wpv,Ma:2023cwe,Ma:2023vvr}. These approaches are often accompanied by concerns regarding overfitting, the stability of the fit, and potential contamination introduced by the filters. The waveform decomposition developed in this work provides an alternative approach with a clear theoretical foundation, enabling a more direct interpretation of the physical content encoded in the waveform. Future extensions could incorporate more realistic source terms, BH spin, and self-force effects, which will be essential for interpreting extreme-mass-ratio inspiral waveforms \cite{Kuchler:2025hwx} and for improving waveform modeling.

Another promising direction is to connect this decomposition with existing approaches to modeling direct waves \cite{Oshita:2025qmn,McWilliams:2018ztb,Kankani:2026byb,Kankani:2025gqj}. In particular, clarifying the relationship between the Green’s function structure and methods such as the steepest-descent approach \cite{Oshita:2025qmn} and the Backwards One Body model \cite{McWilliams:2018ztb,Kankani:2026byb,Kankani:2025gqj} will be important for developing accurate waveform models of direct-wave contributions. Such developments could, in turn, facilitate the extraction of these features from observed GW signals \cite{Lu:2025vol} and enable new tests of General Relativity.

Finally, given that we now have full theoretical control over the prompt response, QNMs, and tail at linear order, it is natural to explore how these components interact at nonlinear order, following the approach developed in \cite{Ma:2024qcv,Khera:2024bjs}. In addition, accurate numerical-relativity techniques, such as Cauchy–characteristic matching \cite{Ma:2024hzq,Ma:2023qjn}, have recently been developed to model these three contributions in waveforms from binary BH mergers. A detailed comparison between these theoretical predictions and numerical-relativity simulations would therefore be both timely and illuminating.

\begin{acknowledgments}
We thank Junquan Su, Neev Khera, Marc Casals, Abhishek Chowdhuri, and Huan Yang for useful discussions.
Research at Perimeter Institute is supported in part by the Government of Canada through the Department of Innovation, Science and Economic Development and by the Province of Ontario through the Ministry of Colleges, Universities, Research Excellence and Security.
\end{acknowledgments}  



\def\bibsection{\section*{References}}
\bibliography{References}

\end{document}